\documentclass[a4paper]{article}

\usepackage{graphicx}
\usepackage{wrapft}

\title{
Preferential attachment scale-free growth model \\ with random fitness
}

\author{ Marcelo D. S. de Meneses $^1$
, Sharon D. da Cunha $^1$
,\\ D. J. B. Soares $^{1,2}$
and L. R. da Silva $^1$
}
\date{}

\begin{document}

\maketitle

\begin{center}
\textit{
$^1$Departamento de F\'{\i}sica Te\'{o}rica e Experimental, \\Universidade
Federal do Rio Grande do Norte, \\Campus Universitario, 59072-970
Natal-RN, Brazil \\
$^2$Departamento de Biof\'{i}sica e Farmacologia, Centro de Bioci\^{e}ncias,
\\Universidade Federal do Rio Grande do Norte,\\
Campus Universit\'{a}rio 59072 970, Natal-RN, Brazil
}
\end{center}

%-------------------------------------------------------------------------

\begin{abstract}
We introduce a model which consists in a planar network which grows by
adding nodes at a distance $r$ from the pre-existing barycenter. Each
new node
position is randomly located through the distribution law
 $P(r)\propto 1/r^{\gamma}$  with $\gamma > 1$.
The new node $j$ is linked to only one pre-existing node
according to the probability law
$P( i\leftrightarrow j)\propto \eta_i k_i / r_{ij}^{\alpha_A}$
($1 \leq i < j$; $\alpha_A \geq 0$);
$k_i$ is the number of links of the $i^{th}$ node, $\eta_i$
is its fitness (or quality factor), and $r_{ij}$ is the distance. We
consider in the present paper two models for
$\eta_i$. In one of them, the {\it single fitness model} (SFM\cite{Soares}),
we consider $\eta_i=1 \; \forall \, i$. In the other one, the {\it
uniformly distributed fitness model} (UDFM), $\eta_i$ is chosen to be
uniformly distributed within the interval $(0,1]$. We have determined
numerically the degree distribution $P(k)$.
 This distribution appears to be well fitted with
 $P(k)= P(1)\, k^{-\lambda}\,e_{q}^{-(k-1) / \kappa}$ with $q \ge 1$,
where $e_{q}^{x} \equiv [1+(1-q)x]^{1/(1-q)}$ ($e_1^x=e^x$)
 is the $q$-exponential function naturally emerging within
 nonextensive statistical mechanics.
 We determine, for both models, the entropic index $q$ as a function of
$\alpha_A$ ($q$ independs from $\gamma$).
 Additionally, we determine the average topological (or chemical)
distance within the network, and the time evolution of the average number of links
$\langle k_i \rangle$. We obtain that, asymptotically, $\langle k_i\rangle\propto(t/i)^{\beta}$,
($i$ coincides with the input-time of the $i$ node) and $\beta(\alpha_A)$ to both cases.
\end{abstract}

%-------------------------------------------------------------------------

Scale-free networks are very popular nowadays \cite{Watts,Bascience,BAmodel,Rozenfeld} due
to their uncountable applications in different fields of knowledge. These models are typically
associated with some physical quantities that are characterized by power-law asymptotic behavior,
instead of the usual exponential laws. Most of these models do not take into account the
node-to-node Euclidean distance, i.e., the geographical distance. One of them which does take
into account this aspect of the problem has been introduced recently \cite{Soares} and discussed.
In this example, as well as in others \cite{Thurner,Natasa}, strong connection has been revealed
with nonextensive statistical mechanics \cite{Tsallis,GellMann}. In the present paper, we follow
the lines of \cite{Soares} which we extend in the sense that we include now a local variable
denominated {\it fitness} or {\it quality factor}, we call this variable $\eta_i$. The model
studied by \cite{Soares} is herein referred to as the {\it Single Fitness Model} (SFM), that is,
$\eta_i=1 \; \forall \, i$, and we also study the {\it Uniformly Distributed Fitness Model} (UDFM),
in which the local fitness is an independent random variable. We are
here interested to focus on the effects of the fitness on the connectivity probability
distribution and similar quantities.

\begin{figure}[!htb]
\centering
\includegraphics[scale=0.15]{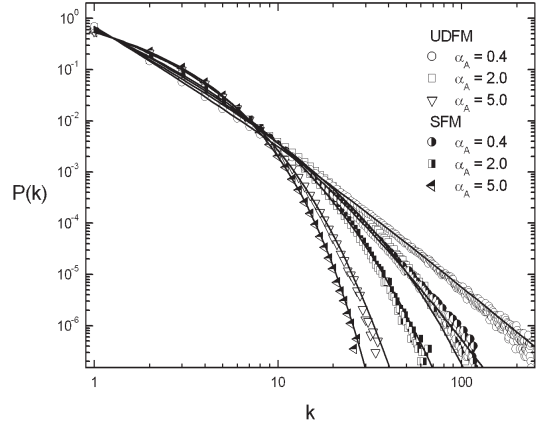}
\caption{Conectivity distribution $P(k)$ in log-log scale for typical values $\alpha_{A}$
for UDFM and SFM models. The symbols are numerical results and continuous
lines are the best fits in according to equation \ref{eq2}. }
\label{fig1}
\end{figure}

\begin{figure}[!htb]
\centering
\includegraphics[scale=0.15]{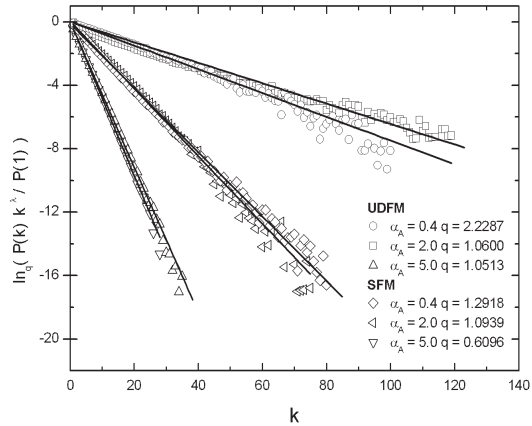}
\caption{Representation of the figure \ref{fig1} in $ln_{q}$ - linear.}
\label{fig2}
\end{figure}

At the plane, we have constructed a complex network which is constituted by nodes and
links between nodes. The network grows sequentially by adding a node at each time. We
fix the first node $(i=1)$ at the origin. The position of the second node ($i=2$)
is chosen randomly with respect to origin in such way its distance $r$ obeys the
probability law $P(r)\propto 1/r^{\gamma}$ with $\gamma > 1$. In this paper,
we have chosen $\gamma=5/4$ for all cases studied. Next the second node is linked to first one.
Then, the origin is moved into network barycenter formed by the two nodes. The previous
procedure is used to include the third site, fourth site, and so on. We link the third
node ($j=3$) to one of the pre-existing nodes in the network. We establish its connection
by considering the linking probability given by $P(i\leftrightarrow j=3)\propto \eta_i \,
k_i / r_{ij}^{\alpha_A}$ ($\alpha_A \geq 0$) and $1 \leq i < j$. The connectivity of $i^{th}$
node is stood for $k_i$ and $\eta_i$ is its fitness and $r_{ij}$ is its distance to new node
$(j=3)$. At this early stage we have  $k_1=1$ and $k_2=1$.

\begin{figure}[!htb]
\centering
\includegraphics[scale=0.10]{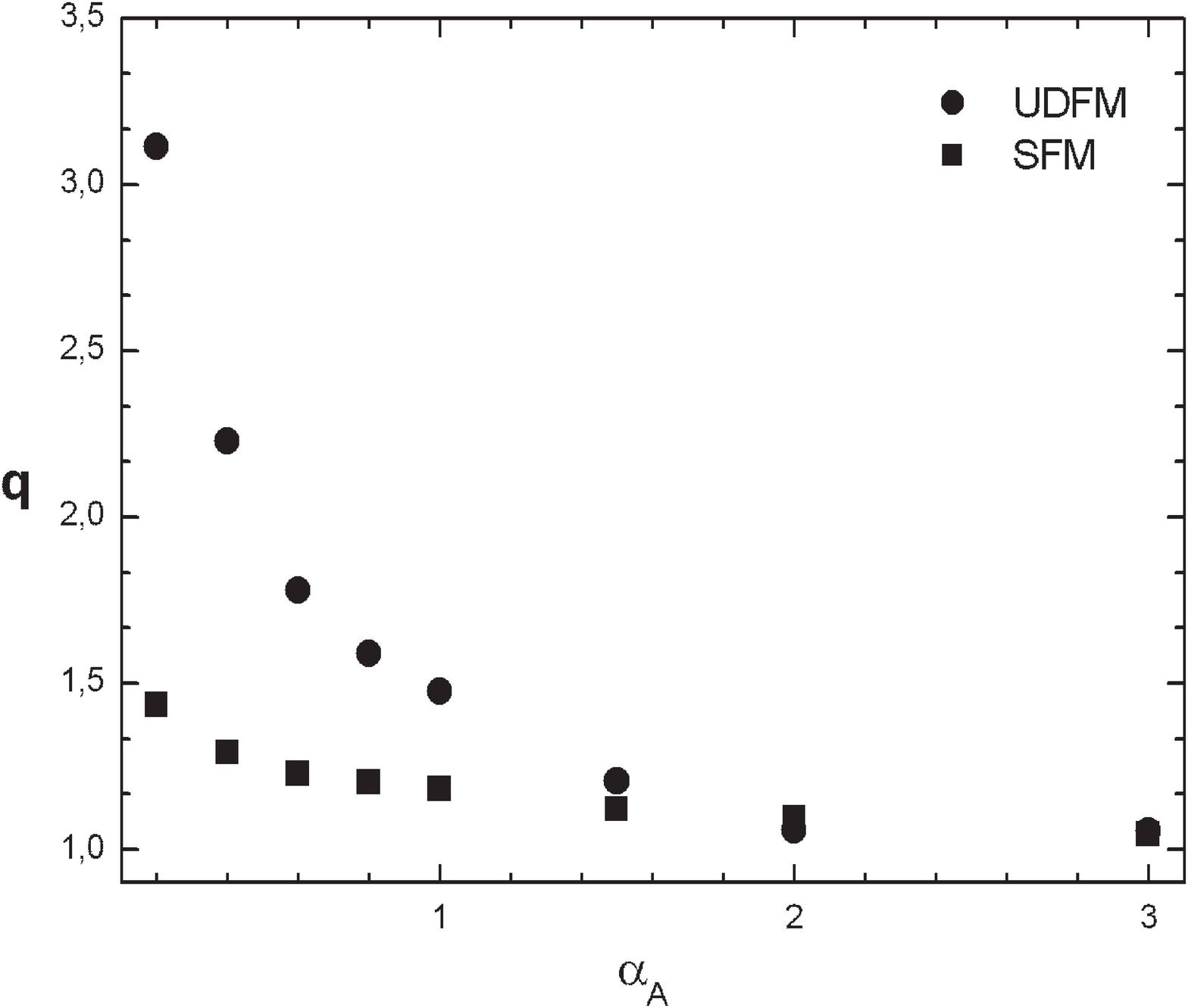}
\includegraphics[scale=0.10]{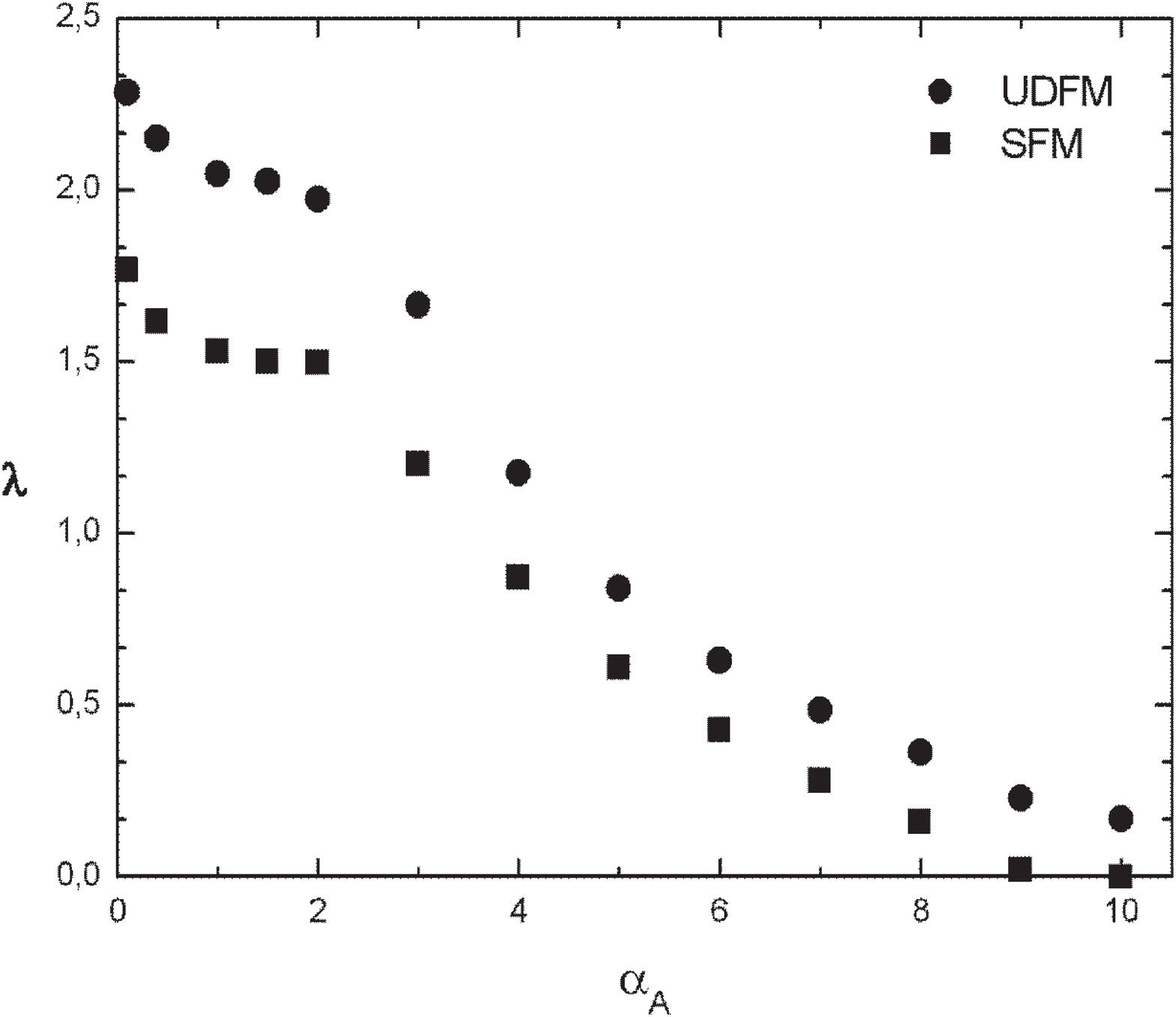}
\includegraphics[scale=0.10]{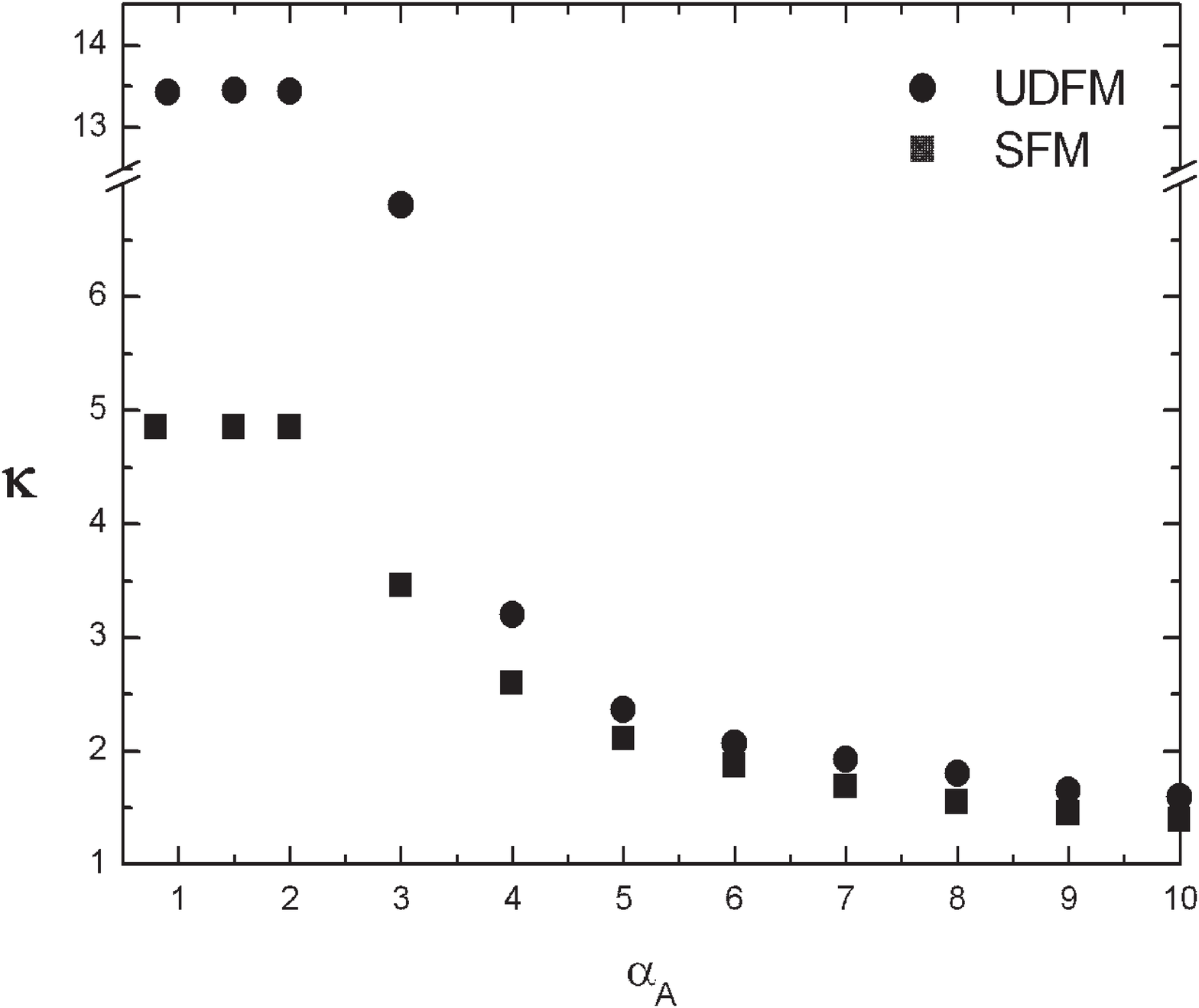}
\caption{$\alpha_A$-dependence of $(q,\, \lambda,\, \kappa)$ for both SFM
and UDFM models.}
\label{fig3}
\end{figure}

The Fitness model is characterized by the presence of factor $\eta_i$
that appear in the linking probability. We define the fitness or quality model
in following way: To each node $i<j$ is assigned a uniform random value $\eta_i \in (0,1]$
when the node is born. That value is kept constant while the
network grows.
In traditional models old nodes are more attractive than new ones. In this
model we give a chance for new sites, with high fitness, become competitive.

The network growth process is sequentially repeated up to size desired
for network. To avoid node position overlap we consider the lowest distance
between them to be unit. If we denote $N$ the total number of nodes one can
write linking probability for $i^{th}$ network node to connect the new node
$j=N+1$ as
\begin{equation}\label{eq1}
P( i \leftrightarrow j=N+1 )=\frac{\eta_i k_i /
r_{i}^{\alpha_A}}{\sum_{i=1}^{N}\eta_i k_i / r_{ij}^{\alpha_A} }
\end{equation}

The dynamics that we have introduced by the above rule does privilege
the connection between new nodes and those having many links (high fitness nodes) since they are
not so far from each other. One similar rule was explored \cite{manna} for the particular case of
uniform distribution of sites within some limited region at the plane, with $\eta_i=1 \; \forall i$.
The $\alpha_A$ parameter control the influence
range of the nodes (or hubs). The case $\alpha_A$ zero corresponds to the
Bianconi fitness model \cite{Bianconi} where distance is not relevant.
In the present paper we will emphasize the following aspects:
\begin{itemize}
\item The network degree distribution in the stationary-state $P(k)$
relative to number of nodes having $k$ links in the $N\rightarrow
\infty$.
\item The temporal dependence of average number of links $\langle k_i
\rangle$
and more precisely to know how asymptotically it increase with time $t/i$, $t\geq i$.
\item The dependence of average length, $\langle l \rangle$, with
$\alpha_A$ parameter.
\end{itemize}

The numerical results are obtained for networks with $10\,000$ nodes
and $2\,000$ runs. In this network size the probability distribution $P(k)$
practically enter in the stationary regime. The figure \ref{fig1} show our
numerical results for $P(k)$ (symbols) and fits (continuous lines) for UDFM
and SFM models. One can observe in our results that $\alpha_A$ has a strong
influence on $P(k)$. The present results are very well fitted by the function
\begin{equation}\label{eq2}
P(k)= P(1) \, k^{-\lambda}e_{q}^{-(k-1)/\kappa}
\end{equation}
where q-exponential function is defined by
\begin{equation}
e_{q}^{x} \equiv [1+(1-q)x]^{1/(1-q)} \;\;\;\;(e_1^x=e^x) \,.
\end{equation}

The variable $\kappa$ is the characteristic number of links  \cite{Soares}.
In the figure \ref{fig2} we represent the figure \ref{fig1} in a $ln_q$ - linear
plot. Where $ln_q(x)=(x^{1-q}-1)/(1-q)$.

\begin{figure}[!htb]
\centering
\includegraphics[scale=0.19]{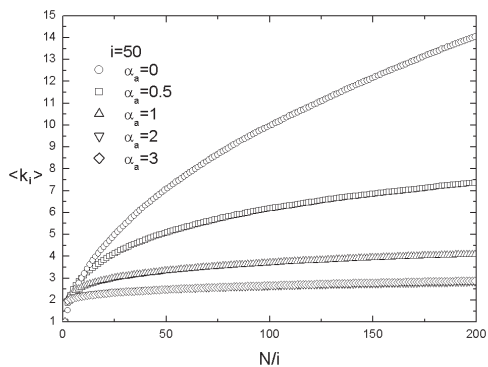}
\caption{Temporal dependency of the average connectivity for UDFM model.}
\label{fig4}
\end{figure}

\begin{figure}
\centering
\includegraphics[scale=0.14]{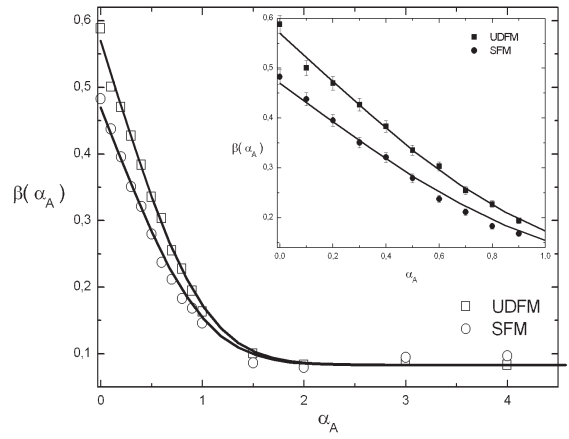}
\caption{Average connectivity exponent for $\alpha_A$ values relative
to measures on node $i=50$.}
\label{fig5}
\end{figure}

\begin{figure}
\centering
\includegraphics[scale=0.14]{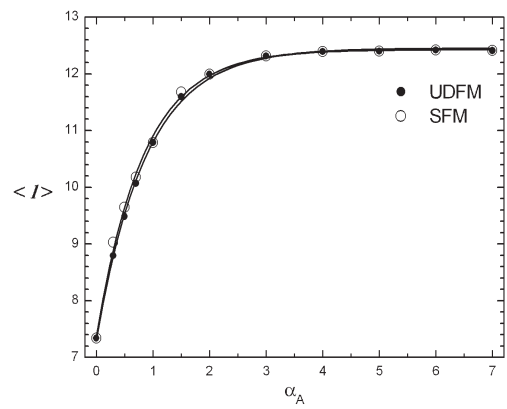}
\caption{Average path length for UDFM and SFM models. }
\label{fig6}
\end{figure}

Analizing the figure \ref{fig1}, we have found that the UDFM has raised the number
of links per node and it turns out that hubs are more connected. In the figure \ref{fig3}
we present the $(q,\,\lambda,\,\kappa)$ evolution curve with $\alpha_A$ for both models studied.
We observe that UDFM curve is always above the SFM case. They have the same asymptotic limit,
i.e., $q(\alpha_A)=1$.

The results for the node average connectivity $\langle k_i \rangle$ are indicated in
the figure \ref{fig4} where we show its temporal evolution. Our data for
$\langle k_i \rangle$ can be fitted by the function
\begin{equation}
\langle k_i \rangle \propto \left(\frac{t}{i}\right)^{\beta( \alpha_A,
\eta_i
)}\;t\geq i
\end{equation}

The collapse of the curves for $\alpha_A \geq 2$ indicates a upper limit
(natural cutoff) for a node connectivity. As much we increase $\alpha_A$ all
hubs reduce its connectivity to a minimal value $2$. We explain this result
imagining that the linking probability decays strongly for increasing $\alpha_A$
in such way that a node does not see other unless its first neighbours,
and therefore they form long node chains. This fact is inherent to the network
growing rule, i.e., if we use other rule for example each new site
links with more than one node, this limit result would be quite different.

In the figure \ref{fig5} we show the exponent $\beta(\alpha_A)$ for UDFM and SFM models.
The exponent decline down from $0.59$ for UDFM and from $0.48$ for SFM to a stationary value $0.1$.
The figure \ref{fig6} shows the average length $\langle l\rangle$ dependency on $\alpha_A$
for UDFM and SFM models. The average path length, $\langle l\rangle$, of a network is
defined as the number of edges in the shortest path between two nodes averaged over all pairs
of nodes. We see that, when $\alpha_A$ varies, $\langle l\rangle$ raises rapidly to a
saturated value $\sim 12$. The saturation in figure \ref{fig6} is explained by the
$\langle k_i \rangle$ convergence in figures \ref{fig5} and \ref{fig4} for $\alpha_A\geq 2$.

Now we will summarize our present paper. Following the lines of the paper \cite{Soares}
we have presented a model in which we add fitness. We study the effect of competition
between the relevant variable for connectedness when we include fitness on the network.
In UDFM model the popular nodes (many links) do compete with younger nodes
when the fitness is a important factor that permits them to get more links.
When we raise $\alpha_A$ we favor the linking between first neighbours in the network.
When  $\alpha_A$ is zero we recover the Bianconi-Barab\'{a}si model \cite{Bianconi}.
The average connectivity $\langle k_i \rangle$ and average path length $\langle l\rangle$ are
appreciably influenced by factor $\alpha_A$ for UDFM and SFM models. We have shown
that degree distribution $P(k)$ is very well fitted by the funtion proposed that contain
$q$-exponential function that emerges naturally from non-extensive statistics. And when
we compare our results relative to SFM we have obtained that the number of links per node has
raised for all values $\alpha_A$. When the $\alpha_A$ is increased we find that
links per hubs has reduced as SFM but it occours slower. The presence of fitness
changes the universality class in the complex network. Then if you want a more
complete scenario for this kind of system, fitness should be included.

%\newpage

\section*{Acknowledgments}

We acknowledge Constantino Tsallis for critical reading and many
fruitful suggestions, as well as financial support from Capes, Pronex/MCT and
CNPq (Brazilian agencies). One of us (LRS) also acknowledges warm
hospitality at the Santa Fe Institute, where this work was completed.


\begin{thebibliography}{20}

\bibitem{Soares}D.J.B. Soares, C. Tsallis, A.M. Mariz and L.R. da Silva,  Europhys. Lett.,\textbf{70},(2005),70.

\bibitem{Watts}D.J. Watts and S.H. Strogatz, Nature,\textbf{393}, (1998), 440.

\bibitem{Bascience}R. Albert and A.-L. Barabasi, Science ,\textbf{286}, (1999), 509.

\bibitem{BAmodel}R. Albert and A.-L. Barabasi, Rev. Mod. Phys.,\textbf{74}, (2002), 47.

%\bibitem{Banature}R. Albert, H. Jeong and A.-L. Barabasi,\JL{ Nature ,401, 1999, 130}.

%\bibitem{nature}H. Jeong, S.P. Mason, Z.N. Oltavai and A.-L.

%Barabasi, Nature ,\textbf{411}, (2001), 41.

\bibitem{Rozenfeld} A.F. Rozenfeld, R. Cohen, D. ben-Avraham and S. Havlin ,Phys. Rev. Lett. \textbf{89},(2002),218701.

\bibitem{Thurner}S. Thurner and C. Tsallis, Europhys. Lett. {\bf 72}, 197 (2005).

\bibitem{Natasa}D.R. White, N. Kejzar, C. Tsallis, D. Farmer and S. White, cond-mat/0508028.

\bibitem{Tsallis}C. Tsallis,  J. Stat. Phys. {\bf 52}, 479 (1988). For full bibliography see http://tsallis.cat.cbpf.br/biblio.htm

\bibitem{GellMann}M. Gell-Mann and C. Tsallis, eds., {\it Nonextensive Entropy - Interdisciplinary Applications} (Oxford University Press, New York, 2004).

%\bibitem{Barabasi} R. Albert and A.-L. Barabasi , \PRL{85,2000,5234}.

\bibitem{manna}  S.S. Manna and P. Sen, Phys. Rev. Lett.{\bf 66 },(2002), 066114.

\bibitem{Bianconi} G. Bianconi and A.-L. Barabasi, Europhys. Lett.,{ \bf 54},(2001),436.

\end{thebibliography}
\end{document}